\documentclass{ws-mpla}
\usepackage[super]{cite}
\usepackage{graphicx}
\begin{document}

\catchline{}{}{}{}{}

\title{Thomas-Fermi Model in Rindler Space}

\author{\footnotesize Sanchita Das$^{a)1}$, Sutapa Ghosh$^{b)2}$ and Somenath Chakrabarty$^{a)3}$}

\address{${a)}$Department of Physics, Visva-Bharati, Santiniketan, 
India 731235\\
$^{b)}$Department of Physics, Barasat Govt. College, Barasat 700124, North $24$'Pgs, India\\
$^1$Email:sanchita.vbphys@gmail.com\\
$^2$Email:neutronstar12@yahoo.co.in\\
$^3$Email:somenath.chakrabarty@visva-bharati.ac.in
}

\maketitle

\pub{Received (Day Month Year)}{Revised (Day Month Year)}

\begin{abstract}
In this article we have investigated the Thomas-Fermi model for the electron gas
in Rindler space. We have observed that if the uniform
acceleration is along $+x$-direction, then there is $y-z$-symmetry in space. For the sake of mathematical
simplicity, we have therefore 
assumed a two dimensional spatial structure ($x-y$) in Rindler space. It has been shown that in two 
dimensional spatial coordinates the electrons are distributed discontinuously but in a periodic manner 
in a number of rectangular strip like domains along $\pm y$-direction. 
Some of them are having void structure, with no electrons 
inside such rectangular strips, while some are filled up with electrons. We call the later type domain as the normal
zone. We have also given physical interpretation for such exotic type electron distribution in Rindler space.
\end{abstract}

\ccode{PACS Nos.: 
03.65.Ge,03.65.Pm,03.30.+p,04.20.-q }

The well known Lorentz transformation in special theory of relativity is  between two inertial frames, i.e., 
having a uniform relative velocity between the frames \cite{R1,R2}. 
A similar type transformation is also possible between an 
inertial frame and a frame undergoing an uniform accelerated 
motion \cite{R3,R3m,R4r,R31,R32,R4,R5,R6,R7,R71,R72}. 
In the later case, it is called the Rindler coordinate transformations. Further in the case of Rindler
transformation, the  
space is called the Rindler space, whereas the former one is the well known Minkowski space. The Rindler
transformations are therefore exactly like Lorentz transformation. The only difference is that the frame is
undergoing an uniform accelerated motion. The Rindler space is therefore also flat like the Minkowski space.

Now from the principle of equivalence, an accelerated frame in absence of gravity
may be replaced by a frame at rest but in 
presence of a gravitational field. The strength of the gravitational field is exactly equal to the
magnitude of the acceleration of the moving frame. Therefore in the case of a frame undergoing an accelerated motion
along positive $x$-direction with an uniform acceleration $\alpha$ is equivalent to a 
frame at rest in presence of a constant gravitational field $\alpha$ \cite{R4r}. 
Then following the references \cite{R4r,R31,R32,R4,R5,R7,R71}, the Rindler coordinate transformations are given by
\begin{eqnarray}
t&=&\left(\frac{1}{\alpha}+x^{'}\right) \sinh(\alpha t^{'})\\
x&=&\left(\frac{1}{\alpha}+x^{'}\right) \cosh(\alpha t^{'})\\
y&=&y^{'}\\
z&=&z^{'}
\end{eqnarray}
where the primed coordinates are in the non-inertial frame. Hence it is trivial to show that the metric 
tensor In $3+1$ dimension is given by (here we have discarded the prime symbols and considered the natural units
with $c=\hbar=1)$
\begin{equation}
g^{\mu\nu}={\rm{diag}}\left((1+\alpha x)^2 ,-1,-1,-1\right)
\end{equation}
whereas in $1+1$ dimension it can be expressed in the following form
\begin{equation}
g^{\mu\nu}={\rm{diag}}\left((1+\alpha x)^2 ,-1\right)
\end{equation}
where $\alpha$ is the uniform acceleration of the moving frame along 
positive $x$-direction, which is also the constant gravitational field along negative $x$-direction for the reference 
frame at rest.
From the Rindler transformations, one can show very easily that the square of the length element
in four dimension is given by
\begin{equation}
ds^2=\left(1+\alpha x\right)^2 dt^2-dx^2-dy^2-dz^2
\end{equation}
This four length element can be shown to be invariant under Rindler transformations. 
Further, assuming the direction of motion with uniform 
acceleration along positive $x$-direction, then we have $dy=dz=0$. 
Using $ds^2$ from eqn.(7) for $1+1$ dimension and following Landau
and Lifshitz \cite{R1}, it can be shown that the single particle Lagrangian  is given by:
\begin{equation}
L=-m_0 c^2\left[\left(1+\alpha x\right)^2 -v^2\right]^{1/2}
\end{equation}
where $m_0$ is the rest mass of the particle and $v$ is the single particle velocity along $x$-direction.
Hence using  the standard technology of classical mechanics, the single particle Hamiltonian in the relativistic
form is given by 
\begin{equation}
H=\left(1+\alpha x\right)\left(p^2 +m_0^2 \right)^{1/2}
\end{equation}
Whereas in the non-relativistic scenario we have
$$
H=(1+\alpha x)\left (\frac{p^2}{2m_0}+m_0\right ) \eqno(9a)
$$
Now it has been shown in the literature that the free Riemann space transforms to a refracting medium 
in presence of strong gravitational field \cite{R1,R8,R9}. Further, if the gravitational  
field changes from point to point then the free space 
will behave like overlapping  refracting media with continuously varying refractive index. 
Stated otherwise, the free space will 
behave like a number of overlapping refracting media with continuously varying refractive index. 
With such a concept of varying refractive 
index, one can very easily explain the phenomena of bending of electromagnetic as well as de Broglie waves 
(matter waves) near strongly gravitating 
objects. This may be treated as an alternative explanation for bending. 
This concept of refracting media in presence of varying gravitational field 
is also applicable in Rindler space. In the later case we have to consider a number of reference frames with various 
values of uniform gravitational field along the same direction.
In other wards these reference frames are assumed to be undergoing accelerated 
motion with different values of uniform accelerations.

Then following Landau and Lifshitz \cite{R1} (see also \cite{R10,R11,R12}) 
it is trivial to show that in the Rindler space 
both the electric permittivity $\epsilon_e$ 
(or dielectric constant) and the magnetic permeability $\mu_m$ are given by
\begin{equation}
\epsilon_e=\mu_m =g_{00}^{-1/2}=\frac{1}{\left(1+\alpha x\right)}
\end{equation}
In the present work we are interested only in the gravity induced electrical permittivity or dielectric constant.
We further assume that the tensor quantity $\epsilon_e$ is diagonal in nature and since the direction of 
gravitational field is  along $x$-axis, we can write
\begin{equation}
\epsilon_e= {\rm{diag}}\left(\epsilon_{xx},\epsilon_{yy},\epsilon_{zz}\right)=
{\rm{diag}}\left(\frac{1}{\left(1+\alpha x\right)},
\gamma, \gamma \right)
\end{equation} 
where $\gamma$ is a constant $\sim 1$. Considering this diagonal form of tensorial structure for $\epsilon_e$, 
the Poisson's equation in CGS Gaussian unit may be written as 
\begin{equation}
\nabla.\vec D =4\pi \rho
\end{equation}
where $\rho=en_e$, the charge density of electrons in the Rindler space and $n_e$ is the electron number density.

Expressing the displacement vector $\vec D$ in terms of electric field vector $\vec{E}$ in the component form, given
by $D_i=\epsilon_{ij}E_j$, with $\epsilon_{ij}$ is diagonal in nature and finally 
writing each component of electric field in terms of electric potential $\phi(x,y,z)$ in the form $E_i=\partial
\phi(\{q_i\})/\partial q_i$, where $q_i=x$, $y$ or $z$, we have
\begin{equation}
\gamma\left(\frac{\partial^2 \phi}{\partial y^2} +\frac{•\partial^2 \phi}{\partial z^2} \right) + 
\frac{\partial}{\partial x} \left(\epsilon_{xx} \frac{\partial \phi}{\partial x}\right)=4\pi en_e
\end{equation}
Absorbing the constant $\gamma$ in $y$ and $z$ derivatives we can write
\begin{equation}
\left( \frac{\partial^2}{\partial y^2}+\frac{\partial^2}{\partial t^2} \right) \phi(x,y,z)+ 
\frac{\partial}{\partial x}\left(\frac{1}{1+\alpha x} \frac{\partial \phi}{\partial x}\right)=4\pi en_e
\end{equation}
In the cylindrical coordinates with axial symmetry, this equation can be written in the form
\begin{equation}
\left( \frac{\partial^2 \phi}{\partial \rho^2}+ \frac{1}{\rho} \frac{\partial \phi}{\partial \rho}\right)+
\frac{\partial}{\partial x}\left(\frac{1}{1+\alpha x} \frac{\partial \phi}{\partial x}\right)=4\pi en_e
\end{equation}
Since the uniform acceleration is along $x$ direction, 
the electron distribution in Rindler space has $y-z$ symmetry. Therefore for the sake of mathematical simplicity, or
to say to get an analytical solution,
we assume the distribution of electrons in two dimension Rindler space. 
The two dimensional space is indicated by $x-y$ coordinates. Our intention in this work is to investigate the
distribution of degenerate electron gas in two-dimensional Rindler space with Thomas-Fermi approximation. To the
best of our knowledge this study has not been done before.

Further, in two dimensional momentum space the electron density $n_e$ is given by
\begin{equation}
n_e= \frac{1}{2\pi} p_{F}^2
\end{equation} 
where $p_F$ is the electron Fermi momentum.
Then the Poisson differential equation (eqn.(15)) in $2$-dimensional Cartesian form may be written as
\begin{equation}
\frac{\partial^2 \phi(x,y)}{\partial y^2}+\frac{\partial}{\partial x}\left(\frac{1}{1+\alpha x} 
\frac{\partial \phi(x,y)}{\partial x}\right)=4\pi en_e
\end{equation}
Now the well known Thomas-Fermi condition is given by \cite{R13,R14}
\begin{equation}
\frac{p_{F}^2}{2m} -e \phi(x,y)=\mu={\rm{constant}}
\end{equation}
Before we proceed further let us now give a physical interpretation of Thomas-Fermi equation (eqn.(18)). When the
system consists of a large number of electrons (here in Rindler space) and is in a stationary state, the value of
$\mu$, which is the total energy of an electron should be same throughout the system. Then the electrons
anywhere in the system do not have an overall tendency to move towards other parts of the system where the single
particle energy value is less. Since for any natural system the tendency is to have minimum energy configuration,
therefore if the total energy is not same throughout the system, then the electrons will try to occupy the spatial
region where the energy is minimum. An instability will therefore grow in this situation.

We would also like to compare eqn.(18) with the eqns.(66)-(68) of the reference \cite{RUFFINI}. In this extremely
interesting and important piece of work the authors have generalized the isothermal Tolman condition and the
constancy of the Klein potential for charge neutral neutron-proton-electron system in $\beta$-equilibrium in compact
neutron stars in presence of gravity. The authors have considered $\sigma-\omega-\rho$ meson exchange type
interactions for the baryons, i.e., for the neutrons and protons. For the electrons, the 
electromagnetic interaction has been
considered. They have given a formalism to solve these equations self-consistently in the general relativistic case.
They have considered Einstein-Maxwell-Thomas-Fermi equations and obtained the Thomas-Fermi conditions for the
constituents of the neutron stars from the constancy of the Klein potentials for each particle species. However, in
our simplified model we have considered only electrons in presence of an electrostatic potentail in Rindler space,
which is basically flat in nature and the gravitational field is assumed to be uniform in some limited spatial
region. The formulation given in \cite{RUFFINI} can be used to investigate the effect of strong quantizing magnetic
field of neutron stars on the charge particles (protons and electrons) when the Landau levels of these particles will
be populated \cite{SASO}. The formalism can be developed for both Rindler space, which is flat in nature and with
uniform gravitational field  in a limited region and also for the general relativistic scenario with Schwarzschild metric.

Now using eqns.(16) and (18), the expression for the electron density can be written as 
\begin{equation}
n_e= \frac{m}{\pi}\left(\mu + e \phi \right)
\end{equation}
Then we have from eqn.(17)
\begin{equation}
\frac{\partial^2 \phi}{\partial y^2}+\frac{\partial}{\partial x}\left(\frac{1}{1+\alpha x}
\frac{\partial \phi}{\partial x}\right)=4 m e\left(\mu + e \phi \right)
\end{equation}
Redefining $\mu + e \phi$ as $\phi$, we have
\begin{equation}
\frac{\partial^2 \phi}{\partial y^2}\rightarrow \frac{1}{e}\frac{\partial^2 \phi}{\partial y^2} 
\end{equation}
and
\begin{equation}
\frac{\partial}{\partial x}\left(\frac{1}{1+\alpha x} \frac{\partial \phi}{\partial x}\right)\rightarrow\frac{1}{e}
\frac{\partial}{\partial x}\left(\frac{1}{1+\alpha x} \frac{\partial \phi}{\partial x}\right)
\end{equation}
Now making the coordinate transformation  $1+\alpha x=u$ and finally redefining $\alpha y\rightarrow y $, we have 
\begin{equation}
\frac{\partial^2 \phi}{\partial y^2} + \frac{\partial}{\partial u}\left(\frac{1}{u} 
\frac{\partial \phi}{\partial u}\right)= \frac{4me^2}{\alpha^2} \phi(x,y)
\end{equation}
Where  $ 4me^2/\alpha^2=a $ is a constant.
Writing the two dimensional potential $\phi(x,y)$ in the separable form, given by
$\phi (x,y)= U(x)Y(y)$
We have the differential equations satisfied by $Y(y)$ and $U(u)$ as:
\begin{equation}
\frac{1}{Y} \frac{d^2 Y}{dy^2}+\delta - \beta a=0
\end{equation}
and 
\begin{equation}
\frac{1}{U} \frac{d}{du} \left(\frac{1}{u} \frac{dU}{du} \right)- \gamma a -\delta=0
\end{equation}
where $\delta$ is a positive constant assumed to be $> \beta a $ and $ \beta + \gamma =1$.
Defining $\nu^2 = \delta - \beta a$, which is greater than zero, otherwise we get unphysical results,
the first differential equation reduces to:
\begin{equation}
\frac{d^2 Y}{dy^2} + \nu^2 Y =0
\end{equation}
with the solution
\begin{equation}
Y=A\cos {\nu y} + B \sin {\nu y}
\end{equation}
where $A$ and $B$ are two real constants.
The other differential equation is given by 
\begin{equation}
\frac{d^2 U}{du^2}- \frac{1}{u^2} \frac{dU}{du} -uU=0
\end{equation} 
where in the above differential equation we have redefined $bu\rightarrow u$, with $b^3= \gamma a + \delta > 0 $
This differential equation is satisfied by the derivative of Airy function $ A_i^{'} (u)$ and is given by \cite{R15}
\begin{equation}
A_i^{'} (u)= - \frac{1}{\pi} \frac{4}{3^{1/2}} K_{2/3} (\zeta)
\end{equation}
Where $\zeta=\frac{2}{3} u^{3/2}$ with $u$ is a real positive variable and $K_\nu (\zeta)$ is the modified Bessel function 
of second kind, given by
\begin{equation}
K_\nu (\zeta)= \frac{\pi^{1/2} (\frac{1}{2} \zeta)^{\nu}}{\Gamma(\nu + \frac{1}{2})} 
\int_1 ^\infty e^{-\zeta t} (t^2 -1)^{\nu - \frac{1}{2}} dt
\end{equation}
Eqns.(29) and (30) are used to evaluate numerically the derivative of Airy function $A_i^\prime(u)$ and thereby
to obtain the potential $U(u)$ as a function of $u$. 

Now before we go to the numerical evaluation of the derivative of Airy function, 
let us analyze the nature of the solutions given by eqns.(27) and (29). 
Since n $2$-dimension $ n_e \propto \phi(x,y)$ and also $n_e\propto p_F^2$, 
the potential $\phi(x,y)=U(u)Y(y)$ should be positive
definite. 
Since the argument of $U(u)$ is positive the derivative of Airy function is negative in nature. Therefore the other part 
which is $Y(y)$ should also be negative, otherwise the space will be forbidden for the electron distribution, 
or in other wards that particular 
spacial region will behave like a void. In fig.(1) we have shown a few such zones in two dimension, including the central 
zone, 
spread  along both positive and negative directions of $y$-axis. 
In this figure we have shown the surface plot of the potential $\phi(u,y)$. The deep lines are for 
$\phi(u,y)=-4.5\times \exp(-3.5x)\times \cos(y)$,
whereas the curves with relatively light impression are for
$\phi(u,y)=-4.5\times \exp(-3.5x)\times \sin(y)$.
For the sake of illustration we have taken $A=B=\nu=1$. Further in the above expressions, the functional 
form of $U(u)$ is obtained by fitting
the numerical data using a search routine.
Let us now consider the solution
\begin{equation}
Y(y)=A \cos(\nu y)
\end{equation}
Since $Y(y)$ is positive in central domain having boundary from  $-\pi/2$  to  $+\pi/2$ along y-axis, the 
central rectangular strip is a void. Along $x$ or $u$ direction, the lower limit is $u_0$ for $x=0$ and the maximum value
of $u$  is fixed by the vanishingly small value of $U(x)$ (is fixed at $10^{-7}$). Next from $ + \pi/2$ to ${+3\pi}/2$, 
$Y(y)$ is negative. As a consequence the overall nature of the potential is positive in nature, which 
allows the presence of 
electrons in this domain and the distribution of electrons is
symmetric about $y=0$. In this way one can get periodically symmetric distribution of 
the voids and normal structures of the domains along $\pm  y$ direction. The non-uniform nature of electron distribution 
in the normal zones along $y$-direction will be determined by the variation of $Y(y)$ with $y$. 
For the other solution along $y$-axis, given by
 \begin{equation}
 Y(y)=B \sin (\nu y)
 \end{equation}
the region from $\nu y=0$ to $ \nu y = \pi$ is not allowed, whereas $\nu y=0$ to $ \nu y = -\pi$ is the normal zone. 
Unlike the previous case the spatial structure along $y$-direction is asymmetric about $y=0$. 
Therefore half of the central 
strip is void, and the other half is the  allowed zone.
Now for the symmetric solution along $y$-axis (solution given by eqn.(31)), the potential $Y(y)$ will vanish for 
\begin{equation}
y_s= \pm \frac{n \pi}{2 \nu}= \pm 
\frac{n \pi}{2(\delta- \beta a)^{1/2}}, 
\end{equation}
where $n$ is an odd integer. Since the parameter $a\sim \alpha^{-2}$, the 
breadth of both the normal zones and the voids will decrease with the increase in the strength of gravitational 
field $\alpha$. In the extreme case, for $\alpha \rightarrow \infty$, 
\begin{equation}
y_s = \pm \frac{n \pi}{2 \delta^{1/2}}
\end{equation}
For the case of asymmetric  solution of $Y(y)$, given by eqn.(32), we have  $ Y=0$ at
\begin{equation}
y_s= \pm \frac{n \pi}{\nu} = \pm \frac{n \pi}{( \delta-a \beta)^{1/2}},
\end{equation}
where $n=0$ or any 
integer number. In this case also the breadth of both physical and un-physical region will decrease with the increase 
in the strength of the gravitational field $\alpha$. Unlike the symmetric solution case, here in the extreme case for 
$\alpha\rightarrow \infty$, we have 
\begin{equation}
y_s= \pm \frac{n \pi}{\delta^{1/2}}
\end{equation}
Along the longitudinal direction also the 
length of the strips decreases with increase in $\alpha$. 

The physical nature of such periodic nature of voids and normal zones is simply because of the minimization of
electrostatic potential energy for the electrons. In the voids the potential $\phi(u,y)$ are negative. The force
acting on the electrons will then be of repulsive in nature. As a consequence the potential energy of the electrons
will be positive, but reverse is the case for the normal zones with $\phi(u,y) >0$, with negative values for
electron potential energy. This phenomena may be compared
with the existence of charges on the outer surface of a hollow metallic charged sphere.

\medskip
\noindent Acknowledgment: We would like to thank Prof. B.K. Talukdar for some valuable discussion.

\begin{figure}[ph]
\centerline{\includegraphics[width=4.0in]{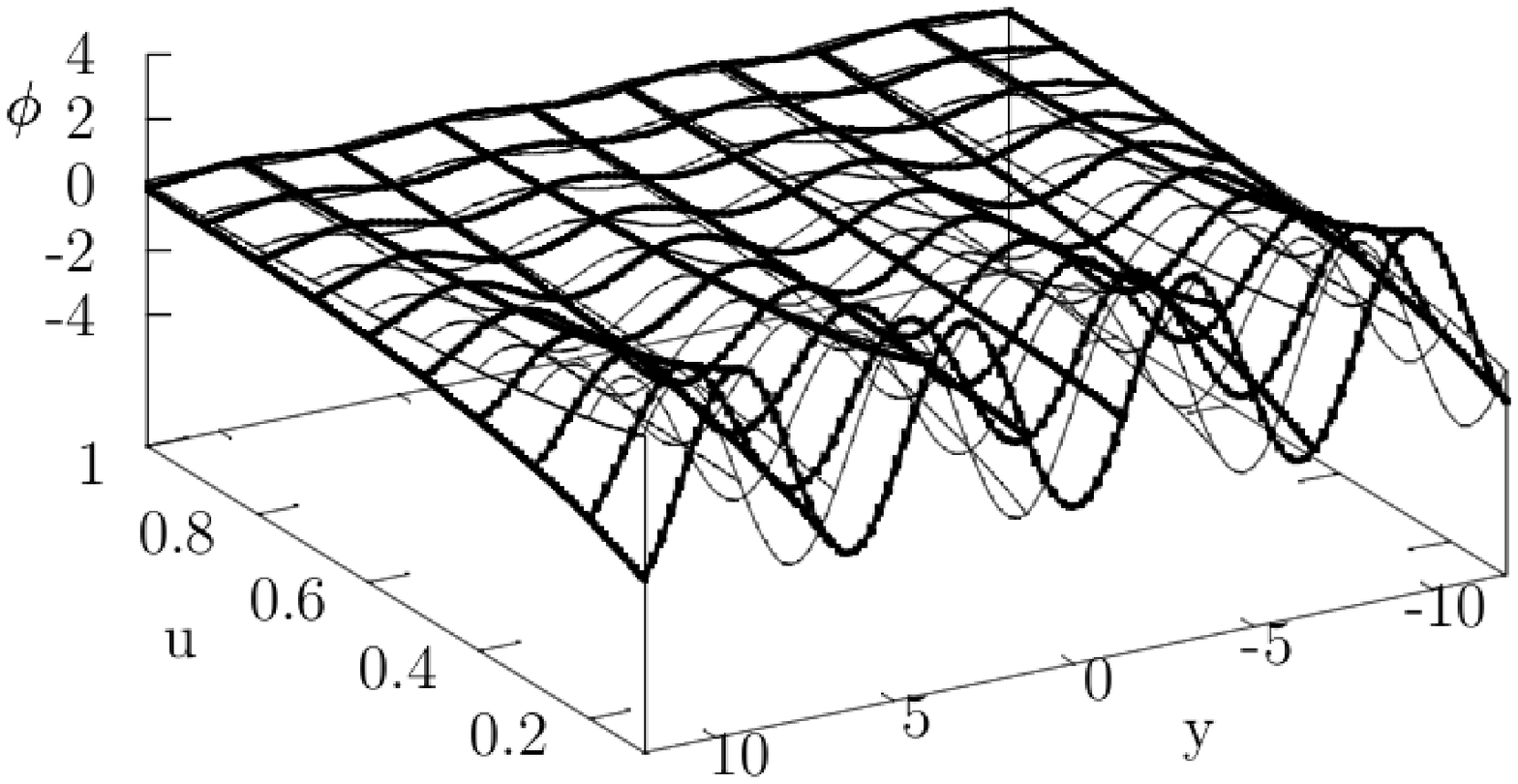}}
\vspace*{8pt}
\caption{Surface plot of the potential
\protect\label{fig1}}
\end{figure}
\end{document}